\documentclass[usenatbib]{ws-jai}
\begin{document}

\catchline{}{}{}{}{} 

\markboth{Willem Baan}{Implementing RFI mitigation in Radio Science}

\title{Implementing RFI mitigation in Radio Science}

\author{Willem A. Baan}

\address{
$^{1}$Netherlands Institute for Radio Astronomy, ASTRON, Dwingeloo, The Netherlands, baan@astron.nl}

\maketitle

\corres{$^{2}$Corresponding author Willem Baan.}

\begin{history}
\accepted{(Published in JAI, VOL. 08, No.01, 2019 )};
\end{history}

\begin{abstract}
This paper presents an overview of methods for mitigating radio frequency interference (RFI) in radio science data. 
The primary purpose of mitigation is to assist observatories to take useful data outside frequency bands allocated to 
the Science Services (RAS and EESS): mitigation should not be needed within Passive bands. 
Mitigation methods may be introduced at a variety of points within the data acquisition system in order to lessen the 
RFI intensity and to limit the damage it does. 
These methods range from proactive methods to change the local RFI environment by means of regulatory manners, to 
pre- and post-detection methods, to various pre-processing methods, and to methods applied at or post-processing. 
\end{abstract}


\section{Introduction}

Radio Frequency Interference (RFI) has become a significant factor in our ability to observe natural phenomena 
on Earth, in our solar system, and the nearby and distant universe. 
From the radio scientist's point of view, RFI is considered to be any unwanted addition to the wanted signal that 
has the potential to degrade or prevent the successful conduct of an observation. 
The chance for encountering RFI has been increasing steadily because of the ability to observe increasingly 
weaker signals with more sensitive detection systems that operate over larger bandwidths.
In addition, because spectral lines are fixed fundamental (rest) frequencies, and radio astronomers are looking at weaker 
emission lines from more distant sources in the universe that are redshifted outside science service bands into bands allocated 
to active radio services.
And while the active spectral usage is intensifying and the used active service bands are moving to higher frequency,  the 
chance for finding strong interfering signals in the science data is becoming larger. 

 Unlike thermal noise, which has stable temporal stochastic properties (white noise) and can be dealt with through radiometric 
 detection (i.e. longer integration times and on-source minus off-source subtraction), an RFI signal is temporally, spatially or 
 spectrally structured and can obscure a wanted signal or produce a false positive detection.
 
The Radio Regulations issued by the Radiocommunication Sector of the International Telecommunication Union (ITU-R) assign Primary and secondary users to allocated spectral bands, such that secondary users are not allowed to interfere with the Primary use of the band \citep{ITU-R}. This regulatory system works well in general, except that some users do not respect these rules.
In 'passive' bands that are allocated to the science services are Primary and all emissions are prohibited. 
However, because this requirement is 'too difficult to adhere to' for neighbouring active services, RFI may need to be tolerated in those bands for 2\% of time for a single network or 5\% for an aggregate situation. 
In addition, one may encounter out-of-band and spurious emissions resulting from emissions from adjacent bands and there may also be some illegal transmissions. 
In bands shared between science services and other services, it depends on the status of the allocated services how RFI issues need to be coordinated.
Outside the bands allocated to the science services, science users have no rights of protection and may operate in these bands on a 'non-interference' basis relative to the Primary users of the band. 


Traditionally much of the data corrupted by RFI has been discarded. However, technology and computing advances have made it possible to mitigate - to lessen in intensity and damage - the effect of the RFI signals in scientific data. 
Yet RFI mitigation has often been an afterthought and retro-active implementation of mitigation options at observatories has been very slow. Often the severity of the anticipated long-term RFI conditions has been underestimated by the users. 
Furthermore, the complexity of integrating RFI mitigation systems into existing observing systems was too high because these systems were not yet fully digital.
Finally, the users preferred (old-fashioned) hands-on flagging methods over the adaptation of black-box procedures.

Changes in technology have made it possible to implement RFI mitigation in existing systems as well as new instrument projects.
Rather than simply complaining about the presence of RFI in the data and about the damage it does, it it possible to do something about the situation.
In this paper we discuss methods of mitigation that have been tried and successfully tested.

Previous reviews covering many of the methods and issues presented in this paper are found in \citep{BriggsK2005, Kesteven2005, BaanB2010,AnEA2018} and ITU-R Report 2126 \citep{Rep2126}

\section{Managing radio frequency interference}

RFI always degrades the quality of the data and the working assumption for most observations has been that RFI-corrupted 
data are partially or completely unusable. 
The most common method for dealing with RFI is to excise spectral or temporal segments of a data set that are known 
to be corrupted.
There are powerful motivations to move beyond throwing away data. Methods that remove or mitigate RFI, thus enabling 
scientific usage of data that would otherwise be discarded, are becoming more and more essential and feasible. 
In addition, the dramatic increase in the size of data sets make automated procedures ever more necessary.

In practical terms and from the point of the science user, any external signal that disturbs the wanted scientific signals can be 
considered as RFI, whether is it strong and showing in the instrument bandpass or weak and only showing in integrated data. 
The ITU-R has defined the 'power flux density' (pfd) levels that are considered detrimental interference that are well within 
the noise of the observing instruments. 
For Radioastronomy they are given in Recommendation ITU-R RA.769 and are based on 10\% of the system noise of the instrument. 
The percentage of permissible data loss in Passive bands resulting from emissions above these thresholds is specified in 
Recommendation ITU-R RA.1513 \citep{ITU-RA}.
In the exclusive Primary bands for the Passive Radio Astronomy Service (RAS) and and the Earth Exploration Service (EESS) are listed in RR No. 5.340, all emissions are prohibited. 
In the other radio astronomy bands, listed in RR No. 5.149, administrations are urged to take all practicable
steps to protect the radio astronomy service from harmful interference \citep{ITU-R}.
For Remote Sensing (RS) and the EESS the interference levels are discussed in 
Recommendations ITU-R RS.1029 for passive sensors, ITU-R RS.1166 for active sensors and ITU-R RS.1263 for 
MetAids \citep{ITU-RS}.
An important Resolution 750(WRC-15) presents compatibility  studies  between EESS(passive) and 
relevant active services and discusses hard limits for OOB emissions into EESS bands \citep{Res750}.

The aim of mitigation efforts is to optimise the increased availability of signal processing hardware and algorithms 
and to enable science observations to be conducted in densely occupied bands and heavily used radio environments.
As a result, RFI mitigation requires more complexity in order to deal with the growing number of wireless 
communications services and the increased need for observations outside protected/allocated bands.

A guiding principle for the implementation of RFI Mitigation should be to remove interfering signals as early as 
possible in the data chain and at the highest possible time and spectral resolution. 
Since the removal of RFI signals is generally limited by the instantaneous signal-to-noise of the signal in the system, 
the strongest and most damaging RFI signals may be mitigated at the sampling speed in the system.
Weaker signals that are still buried in the sampling noise can only be removed after time-integration later in the signal path.
Mitigation of the RFI as early as possible in the data chain minimises the damage to the data and 
keeps the data loss at a minimum level. 

Mitigation at high time resolution is particularly beneficial for time-variable RFI and will reduce data loss due to time-smearing.
Higher spectral resolution reduces the occupation of spectral channels by the RFI and limits the damage of RFI 
to the data due to spectral-smearing. 

In the following sections, methods of mitigation are discussed for a more general structure of receiving systems. 
Most of this discussion is geared towards radio astronomy instruments but many methodologies will also be applicable 
for EESS systems for both terrestrial and space applications. However, specific applications for space-based Remote Sensing applications will not be discussed in this review.

\begin{figure}
\begin{center}
\includegraphics[width= 10cm]{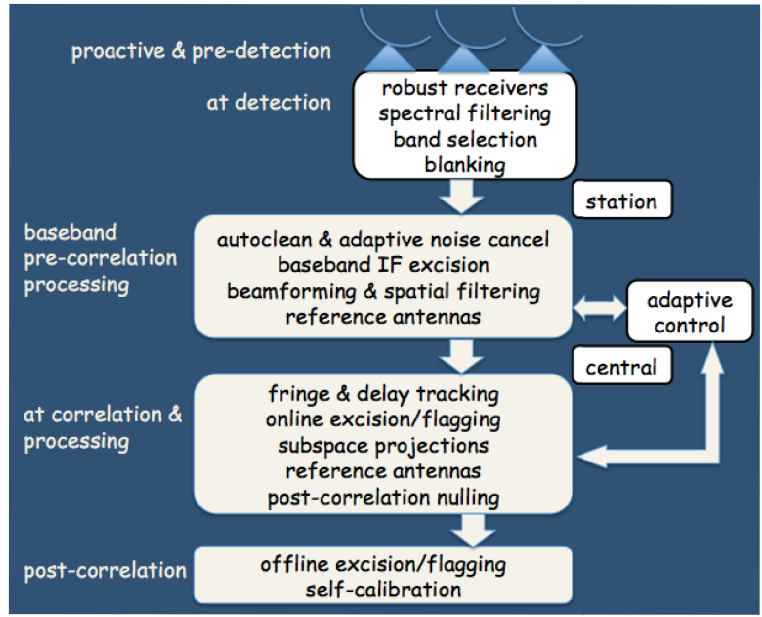}
\end{center}
\caption{Stages and methods of RFI Mitigation along the signal path in a radio observing instrument.}
\label{fig:fig1}
\end{figure}

\section{Implementation in observing systems - a multi-layer strategy} 
 
Effective mitigation of RFI signals needs to be multi-layered using different technical methods at different locations 
in the signal path of the system and can be subdivided in three types of activity:
1) Preventing radio frequency interference (RFI) signals from entering the science data,
including both the reduction of the observatory's vulnerability to RFI signals and the elimination of in-house signals. 
 2) Removing the strongest and/or time-variable RFI signals from the data in real time at one or more locations in the data chain.
 3) Off-line removal or reduction of the impact of RFI.
 
Multi-layer mitigation can be introduced at different stages in the data path and can be grouped into different 
categories (Figure \ref{fig:fig1}).  Naturally, the ability to apply these mitigation protocols will depend on the structure of observing 
system, and some methods cannot be applied to all systems. 
Certain methods can be applied at and before each individual observing station, others in the signal chain of the stations before 
station data processing or before array central processing, others apply to the individual/central processing site, and lastly 
some methods may be applied after the data has been processed.
The ability of all methods to excise or subtract RFI signals from the incoming data is limited by the Interference to Noise Ratio (INR)
and they can reduce the RFI to a level of INR $\approx$1. Therefore strong RFI signals can be dealt with early in the data path, 
while weaker RFI signals can only be addresses after integration.

The multi-layer mitigation scheme presented in Figure \ref{fig:fig1} includes an "adaptive control' component. 
After a 'training process' this unit should be able to recognize the specific type of 
interference encountered and adjust the system to invoke the optimal mitigation algorithm. 
The long-term goal for this unit is to install an 
autonomous RFI mitigation system that can handle all kinds of interference environments.
Another function to be integrated into this control unit is the accurate bookkeeping on 'what has been done where'.

\section{Station - Pro-active and pre-detection} 

{\it Cleaning the Site} - Pro-active and pre-detection actions are aimed at preventing RFI signals from entering the 
detector systems or reducing their strength. 
In the first place, locally generated RFI from inside and outside the observatory needs to be eliminated or reduced to 
levels that are below RA.769 values at the entry of the sensor or telescope. 
This form of Electromagnetic Compatibility (EMS) should address all radio frequency noise generated by electrical and 
electronic equipment and requires shielding/screening or placement in Faraday cages. This effort would also include laptops 
and mobile communication devices in the building and on the site.
In the case of distributed buildings at an observatory site, also the radio noise generated in such buildings should be reduced 
to RA.769 levels at the location of the telescope or sensor.
The observatory should not be its own (worst) enemy.

\vspace{2mm}{\it \bf Quiet and Coordination Zones} - Working close with governmental spectrum and regulatory agencies 
it is necessary to establish a Quiet Zone (QZ) around the observatory.  
It is generally not possible to make such a quiet zone actually radio quiet, but it will prevent new transmitters to be located 
inside the zone and help remove some old ones.
In addition, a Coordination Zone (CZ) should be established that facilitates coordination with the operators about
nearby transmitters in order to reduce or eliminate the signal strength in the direction towards the observatory.
Many radio astronomy observatories have established quiet/coordination zones.
Large-scale zones have been documented for several sites, such as the Mid West Radio 
Quiet Zone in Western Australia \citep{ARQZWA2007}, the National Radio Quiet Zone around Green Bank
\citep{NRQZ1958} and the Puerto Rico Coordination Zone around the Arecibo Observatory \citep{PRCZ1998}.
Information about the structure and characteristics have been described in ITU-R Report RA.2259.
Quiet zones for Earth exploration systems are realistic for terrestrial applications. 
However, RS space applications will mostly depend on coordination with terrestrial interferers.

Transmitters existing inside a Quiet Zone at the time of the establishment of the zone or after need 
to be identified and need to be quieted or relocated with the help of the national spectrum agency.
Friendly coordination with transmitting stations existing in a Coordination zone may result in lowering their power or 
changing their emission pattern or change channel in order to reduce the signal strength at the observing frequencies 
of the observatory. In general, the operators of existing stations are sensitive to the needs of an observatory.
New stations in a coordination zone need to be coordinated before the installation takes place.

\section{Station Mitigation at the detection system}
Mitigation at the detection system should be applied to standalone stations as well as to all stations that are part of an array.
The use of robust receivers for observations that can withstand the presence and effects of strong RFI is essential. 
The suppression of second-order intermodulation products and enough 'headroom' to avoid gain suppression are 
guiding principles for the construction of such receivers.

Spectral filtering and band-edge filtering have been a standard method to avoid strong signals entering the band. 
However, filtering comes at a price in gain performance. For in-band filtering, super-conducting notch filters have been 
developed. These costly devices will slightly degrade the performance of the receiver but their tuning is not very flexible. 

Edge filtering or band selection to avoid band sections with strong signals may be applied for continuum studies, but 
observers enjoy less flexibility with regard to spectral line studies and redshifted surveys. Application of software 
defined radio techniques will aid in finding clean spectrum and avoid RFI signals.

\section{Station Mitigation before correlation and processing}

\subsection{Excision at baseband}

\vspace{2mm}{\it \bf Time and frequency analysis} - 
A powerful way of removing strong RFI signals that have entered the receiver is real-time excision at the Intermediate 
Frequency (IF) (at baseband) at the sampling speed. 
Signals that have significant SNR at the sampling frequency are also most damaging to the surrounding spectral data and 
should be removed from the data before further processing.  
Successful applications of excising RFI from the data in the IF stage have been based on thresholding and kurtosis schemes.  
Real-time thresholding simply applies a threshold of $n$ times the root mean square (rms) of the noise for removing IF signals 
above this level in the frequency and/or time domains. 
A successful implementation of this and other methods has been documented for the Westerbork Synthesis Radio Telescope \citep[WSRT;][]{BaanFM2004,BaanEA2010}.
Recently, a real-time application of the Median Absolute Deviation (MAD) estimator \citep{Fridman2008} for RFI thresholding 
has been implemented (pre-correlation) within the GMRT wideband backend in order to target  time-variable power-line RFI 
particularly at frequencies less than 700 MHz \citep{BuchEA2018}.

Similarly a dynamical excision threshold may be based on a {\it Kurtosis analysis} in order to identify and excise statistical outliers.
Kurtosis is a measure of the 'tailedness' of the probability distribution using a scaled version of the fourth moment of the data such 
that a higher kurtosis number results from infrequent extreme deviations.
This statistical method has been applied successfully for removing signals with kurtosis $> 3$ from solar data \citep{NitaEA2007,GaryLN2010}.
Kurtosis excision at baseband is also applied to pulsar observations at the Parkes telescope \citep{Hobbs2018} and works 
well for RFI signals that are significantly stronger than the pulsed signals of the pulsar. 
Caution should be applied for pulsar searching observations because the pulses for these sources are still unknown.

\begin{figure*}
\begin{center}
\includegraphics[scale=0.45]{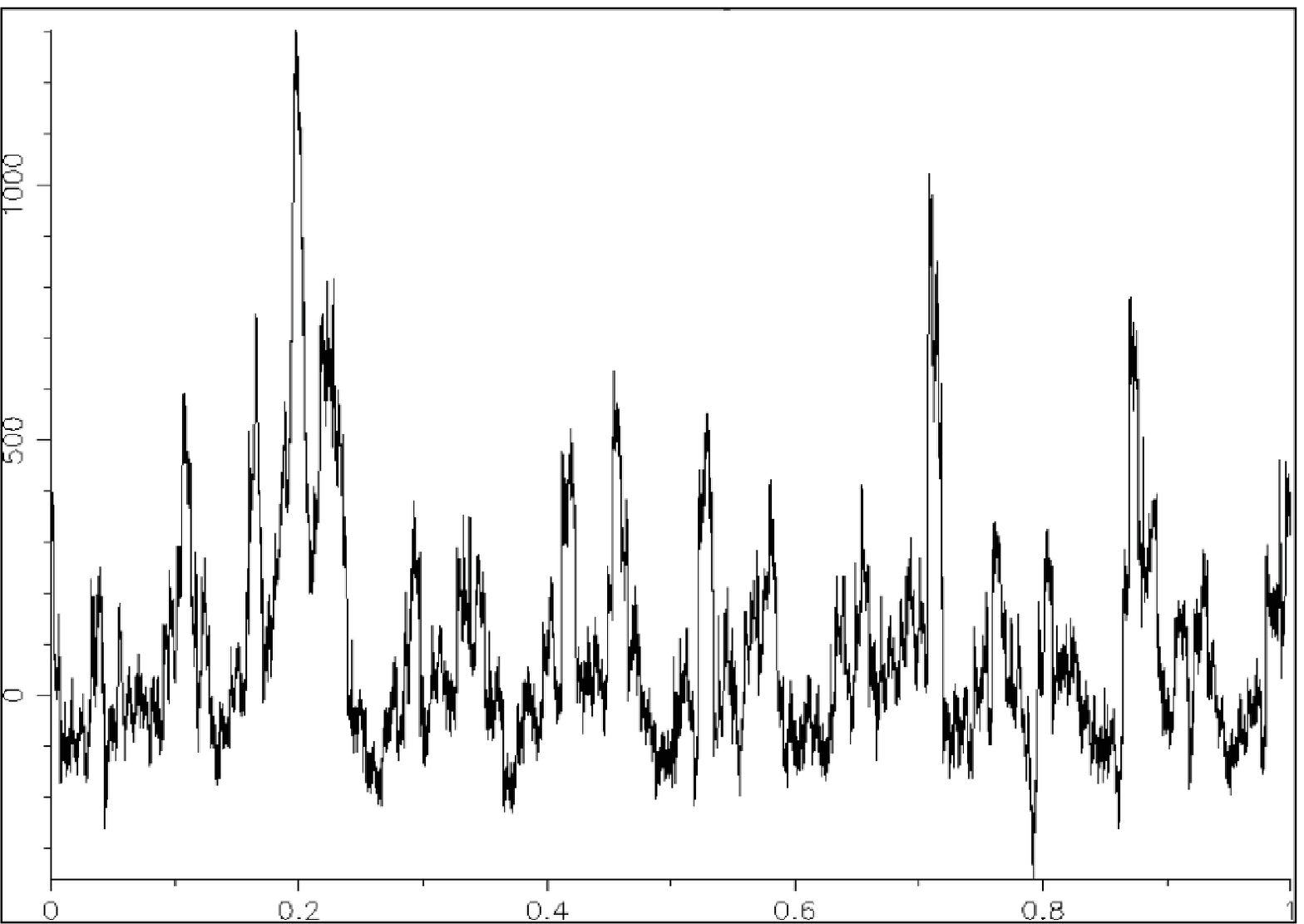}
\includegraphics[scale=0.45]{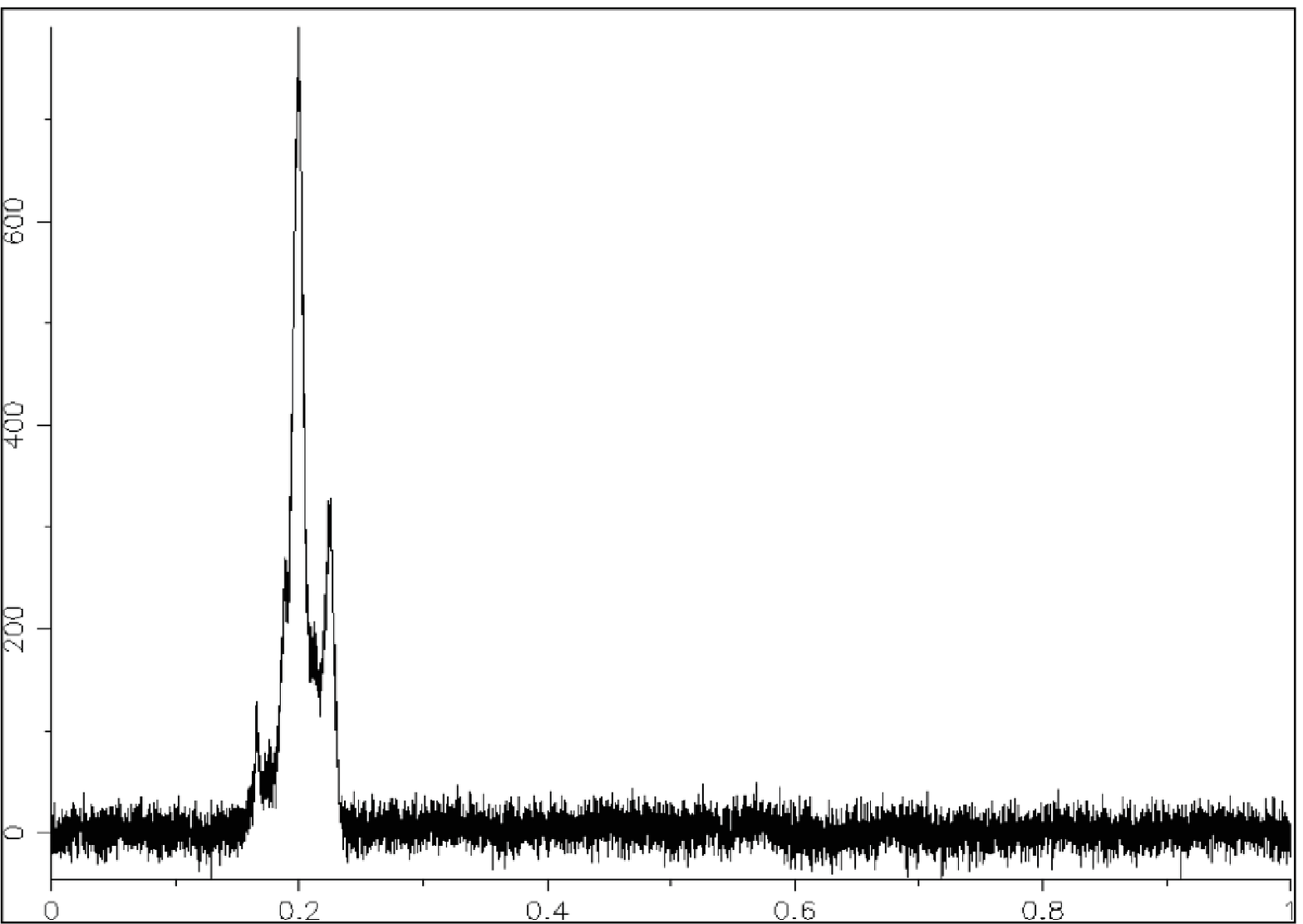}
\end{center}
\caption{WSRT observations of pulsar PSR\,0332+5434 under severe RFI conditions at 1625 MHz. (Left) A period folded 
spectrum without RFI mitigation. (Right) The period folded spectrum with thresholding excision clearly shows the pulsar profile. 
Figures are courtesy of Petr Fridman and Ben Stappers. }
\label{fig:fig2}
\end{figure*}

In particular, time-variable interference can be addressed optimally by these methods, although they will result in a certain percentage of data loss that depends on the bandwidth and duty cycle of the RFI signal. 
However, not removing these signals would result in even more damage and data loss after integration at a later stage. 
As an example, the out-of-band emissions from the uplink-downlink of a Mobile Satellite System with a 50\% duty cycle would 
result in data loss in the adjacent RAS 1612 MHz band in the range range from 50-100\%. 
Real-time thresholding at baseband with the RFIMS observing system at Westerbork Observatory made 
it possible to only sample the uplink time segment of Iridium with good results, but still with a data loss of about 50\% (Figure \ref{fig:fig2}).
For pulsar observations with the new Apertif {\it phased-array feed} (PAF) system at WSRT, thresholding in time and frequency 
is applied on one second integrated samples after the beamforming process together with coincidence detection for all beams 
\citep{vLeeuwen2018}.

Thresholding and kurtosis analysis may be used to clip the power of strong RFI signals and replace these with some 
smaller values. This 'data loss' changes the noise properties in the affected data and may also affect the data contained wanted 
signals in (sometimes) unknown locations in the spectrum. Accurate bookkeeping of the 'when and where' is therefore an 
important part of the process.

Implementation of mitigation and excision at baseband requires the (simple) inclusion of processing power in the IF path of the 
instrument before further processing or correlation. For IF processing in PAFs the compute power required to process all elements 
will be significant, and processing after beamforming can suffice if the RFI does not affect the beamforming process.

\vspace{2mm}{\it \bf Time Blanking} - In the presence of strong periodic signals such as radar systems, a blanking scheme 
may be adopted where the backend and data taking is interrupted starting at the leading edge (in time) of the radar pulse. 
The blanking window will have to take into account both the pulse and the multi-path signals caused by the intervening terrain.
This method has been applied at the Arecibo Observatory for certain radars at the San Juan (SJU) 
airport, which allows observations inside the radar band with a certain data loss that would not have been possible otherwise. 
Thresholding at baseband (discussed below) would achieve the same results as blanking the radar at the backend and with a similar amount of data loss.

\begin{figure}[h]
\begin{center}
\includegraphics[width= 8cm]{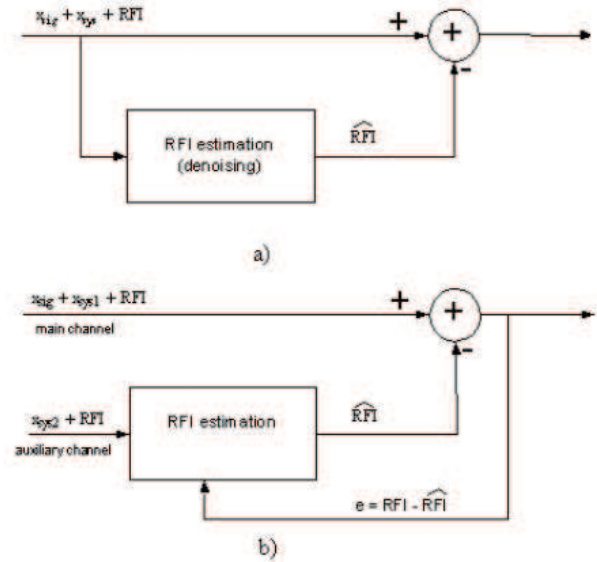}
\end{center}
\caption{Methods of RFI cancellation using autoclean filtering (frame a) and RFI excision by filtering
using a reference channel (frame b). Image obtained from Fridman \& Baan (2001).}
\label{fig:fig3}
\end{figure}

\subsection{Removing the RFI signature}

A number of {\it waveform subtraction} methods and methods that employ the statistical properties of the data have 
been used to cancel and remove the added RFI power from the spectral data in the time domain with minimal or 
no damage to the data itself.  However, the effectiveness and the achieved suppression are limited by the quality of 
the estimate of the interference received by the instrument.

\vspace{2mm}{\it \bf Autoclean} - Temporally spread and strongly correlated RFI can be suppressed using cancellation techniques 
based on estimating the RFI waveform and subtracting it from the {\it signal + RFI} mixture.
The RFI waveform can be estimated using available filtering technique (spine-smoothing, wavelet analysis, Wiener filtering, 
parametric estimation). In this {\it autoclean procedure} the estimate is then subtracted from the input data in the temporal or 
frequency domain (see Fig. \ref{fig:fig3}a). In principle, these methods do little or no damage to the data and the 
signal of interest.

An application of the autoclean procedure has been described employing a WSRT telescope and comparing the result with 
that of an 'uncleaned' antenna \citep{FridmanB2001}. 
An example of a parametric approach to this type of RFI cancellation has been presented for the strong 
interfering signal of a GLONASS satellite operating within the OH 1612 MHz band \citep{EllingsonEA2001}. 
The RFI parameters Doppler frequency, 
phase code and complex amplitude were determined for each data segment, to be used to calculate the RFI waveform for 
subtraction from the {\it RFI + noise signal} mixture. 
A recent parametric application is described for the removal of a broadband digital video broadcast (DVB) signal from the data \citep{SteebDW2018}.

\vspace{2mm}{\it \bf Adaptive Noise Cancellation and Reference antennas} - 
A separate dedicated reference channel may be used to obtain an independent estimate of the RFI signal. 
This technique of {\it adaptive noise cancellation} (ANC) is used actively in digital processing for communication and military 
technology.
Two data channels as shown in the block diagram of Figure \ref{fig:fig3}b are a 'main channel' radio telescope and an 
'auxiliary or reference channel' radio telescope both pointing 'on-source' and containing the RFI signal.
Since the RFI signal in both channels is not exactly identical because of different propagation paths and radio receivers, 
an adaptive filter is used to reduce the error signal and apply the result to the main channel. 
This basic principle of this procedure both in the temporal domain ({\it adaptive filtering}) and also the frequency domain 
is to make an fast Fourier transform (FFT) from the incoming data, perform an adaptation operation on the frequency bins, 
and then return to the 
frequency domain via an inverse FFT. This method, based on Wiener filtering, works for interfering signals with a
significant INR, i.e. when the RFI dominates the system noise, and the suppression of the interfering signal can be 
equal to its instantaneous INR. Examples of the use of reference channels are described in \citep{FridmanB2001}.

A variation on adaptive filtering is to subtract a reference data-channel from a signal data-channel by comparing {\it on-source 
plus RFI} and {\it off-source plus RFI} signals. Separate reference antennas pointed at the interfering source have been 
successfully used at the GBT and at Parkes \citep{BarnbaumB1998,BriggsBK2000,NigraEA2010}. 

Adaptive filters are effective for spectral line observations, where the RFI and the signal-of-interest occupy the same 
frequency domain, and when spectral information is unimportant, such as in pulsar \citep{Kesteven2005} 
and continuum studies. 
However, these methods should preferable be applied before data processing but may also be applied after depending 
on the SNR of the external signal.
In principle, the signatures for multiple interfering sources may be added before subtraction from the wanted signal. 

\vspace{2mm}{\it \bf Higher-order statistics and Probability Analysis} -   
The voltages of the system noise and a radio source generally have a Gaussian signal distribution with a zero mean. 
Fourier transformation of such an ideal signal also gives real and imaginary components in every spectral bin, that are Gaussian random values with zero means. 
On the other hand, the instantaneous power spectrum (the square of the magnitude of the complex spectrum) has an exponential 
distribution that is described as a chi-squared distribution with two degrees of freedom.
The presence of an RFI signal modifies the ideal input signal and yields a change of its statistics, giving a power spectrum with
 a non-central chi squared distribution with two degrees of freedom. Real-time analysis and DSP processing of 
 the distribution before any averaging will allow a separation of the two signal components such as an RFI signal superposed 
 on a known spectral line \citep{FridmanB2001}.

Higher-order statistics methods and determining the moments of the probability distribution of the signal power can be 
used both for spectral line and continuum observations, and it will be useful to introduce this procedure into new detection 
systems \citep{Fridman2001}.
However, reliable estimates of these higher moments (or cumulants) will require more averaging than for the first moment (the mean), which is necessary for mitigation of the weak RFI. On the other hand, large averaging intervals will
smooth the variability of the RFI and will yield estimates with a considerable bias. 
Therefore, there are limitations on the detection and excision of weak RFI signals, which is equally true for all other RFI 
mitigation methods. 


\vspace{2mm}{\it \bf Multi-Element and Phased-Array Systems} - 
In addition to station processing using thresholding or kurtosis techniques, adaptive filtering is particularly useful for 
multi-feed single-dish radio telescopes and array instruments, where one array 
element is being used as a reference channel. For an array instrument each strong and distinct source of RFI requires a 
separate reference antenna.

{\it Spatial filtering} is widely used in {\it smart antennas} in (multi-element) radar and communication systems.
The applicability of spatial filtering for sparse arrays may be limited because of their sparseness and because of offline processing of 
long integrations during which RFI sources may also be moving. 
A variety of specific algorithms including maximum
SNR, subspace projection, Wiener filtering, and multiple sidelobe cancelling have been studied for application to radio astronomical observing \citep{Boonstra2005, BoonstraT2005,Ellingson2003,EllingsonH2002,JeffsEA2005,FridmanB2001,LeshemVB2000, vdTolV2005}.

Multi-layer real-time {\it adaptive/spatial-nulling} techniques may be applied for telescopes with phased-array technology 
but also for densely populated sensor stations, such as for recent low-frequency systems or new generation instruments \citep{vArdenneSH2000,BentumEA2008,BlackEA2015,Boonstra2005,Bregman2000}. 
The use of a reference antenna with a direct look at the interferer with a higher SNR will greatly improve the nulling performance \citep{BriggsBK2000,JeffsEA2005,SardarabadiEA2016}

In order to fully exploit these techniques, multi-element systems should have computer control of the antenna phases 
and their amplitudes \citep{FridmanB2001}. Adaptive filtering using a beam-forming algorithm requires a high INR and 
is limited to a small number of RFI targets to be tracked during an observation. The RFI sources also need to
remain stable and predictable through an observation. Spatial filtering in beam-forming mode for a
limited number of RFI sources generally does not degrade the image generated by the main beam.
 
\vspace{2mm}{\it \bf Subspace Projections for Multi-Element Systems} - 
{\it Subspace projection} for array null-formation identifies the interference in terms of correlations between 
array elements, which can be used to determine beamforming coefficients that result in patterns that reject the 
interference with little or no effect on the main lobe characteristics. 
Subspace projection using outstanding RFI properties has significant advantages for radio astronomy \citep{RazaEA2002} 
but does not help with poor detection and localisation performance for the interference \cite{EllingsonH2002, LeshemVB2000}. 

Beam pattern distortions when subspace nulling a (rapidly) moving source of interference and for narrowband signals 
can be reduced with deeper nulls when time-integrated data is stored and processed  \citep{JeffsW2008, LandonJW2011}.
However, the cost of digital/transport infrastructure and a possible auxiliary antenna is an obstacle for implementation 
and there are limits in nulling below the noise floor \citep{JeffsW2013}. There is also the need for a good science case.
In general, null-forming is most applicable to mitigation of RFI from satellites, and can be expected
to be somewhat less effective against terrestrial RFI with intervening terrain. 

\section{Station and Central - Mitigation at correlation and post-correlation}

As part of the correlation process, digitised data are generally integrated over time intervals ranging
from the sampling time up to seconds, which significantly raises the INR. In consequence,
persistent but weak RFI, that could not be treated in real-time, and weak (spectral) remnants of
earlier mitigation operations become accessible for processing. 
Therefore, a second layer of (automatic) real-time excision can be applied on the accumulated data 
records that would be complementary to off-line flagging of the data. 

Flagging and excising interference in the frequency or time domain has been the standard procedure used by 
radio scientists during post-correlation processing. 
Flagging of baselines and antennas in array observations also identifies and eliminates system problems.
Post-correlation excising is performed on integrated/averaged and correlated data, and can result in considerable 
data loss because even for time- and frequency-variable RFI whole time-slots, whole baselines, and/or whole 
antennas need to be flagged. 
This differs from antenna-based flagging/excising of IF baseband data, which results in less data loss overall.
 
\vspace{2mm}{\it \bf On-line or off-line processing} - Anti-coincidence protocols may be incorporated at the processing 
stage in order to identify 
the RFI components, as well as digital mitigation processing and the integration of a reference antenna during 
(software) correlation. However, the implementation of these algorithms into pre-existing hardware backends requires the 
addition of both special hardware and software.

Automated flagging and excision of calibrated/processed and integrated data records has been proposed and implemented for 
single-dish systems or for each baseline of an interferometry system 
\citep{Kalberla2010,KeatingEA2010,Middelberg2006, OffringaEA2010, SirothiaEA2009}. 
Recent developments on Recurrent Neural Network algorithms and Deep Learning systems may provide different options 
for recognizing/identifying RFI in the data \citep{AkeretEA2017,BurdEA2018}.
Systems incorporating the SumThreshold algorithm have been implemented for the low-frequency LOFAR observatory 
and higher frequency MERLIN interferometer \citep{OffringaEA2012,PeckF2013}.
New generation software correlators permit the integration of threshold or kurtosis-based
flagging applications before and after FX (Fourier Transform before multiplication) correlation and
stacking protocols \citep{Deller2010}. 
A reference antenna implemented at the post-correlation stage can remove the signal from a well-defined 
RFI source using the available closure relations \citep{BriggsBK2000}.

\vspace{2mm}{\it \bf Fringe stopping and delay compensation} - Array instruments employ fringe-stopping and delay-compensation techniques to keep a zero fringe
rate at the central observing position during observations. As a result the stationary (terrestrial) and
satellite RFI components in data distinguish themselves by fringing faster than components from
astronomical sources. This distinctive (relative) motion allows the off-line identification and
elimination of stationary RFI sources from both the correlated data and the image plane without
causing data loss \citep{CornwellEA2004,WijnholdsEA2004}. The coding for this
operation originating at the GMRT is now incorporated into AIPS task UVRFI \citep{Athreya2009}.

\vspace{2mm}{\it \bf Sub-space processing} - 
In addition, more sophisticated statistical or sub-space processing can be implemented to remove the 
RFI component with a minimum of data loss.
{\it Subspace filtering methods} may also be implemented in a digital correlation system to search for a particular signature 
in the RFI power component of data in order to identify and remove it. 
A particularly successful application is the search for cyclo-stationarity within the data, which works well
for digitally modulated RFI signals \citep{FeliachiEA2009, FeliachiEA2010,WeberEA2007}.
For array applications these methods depend on the estimation of the RFI spatial signature using the diagonalization of 
the correlation matrix or the cyclic correlation matrix of the array \citep{HellbourgEA2012}.
In the case of a reference antenna, the evolution with time for a multiple RFI scenario requires a subspace tracking 
approach using the covariance data where the reference antenna supports a faster convergence \citep{HellbourgEA2014}.

\section{Conclusions} 
Both on-line and off-line data processing has been successful in mitigating the RFI environment of radio astronomy 
observatories.  
While there are an increasing variety of viable mitigation options, the choice of method depends 
strongly on the RFI characteristics, the type of radio instrument, and the type of observation. 
In particular, on-line real-time data processing may be preferred for a variable RFI environment, while special 
measures such as reference antennas and spatial filtering may be preferred for known and fixed sources of RFI. 
In addition to these factors, the absence of human involvement may also render automated on-line processing a 
more attractive option. 
 
 No universal method exists for mitigating RFI in astronomical data and no method can identify or remove RFI within 
 the noise of the system. 
 The effective suppression of RFI depends on the INR and its temporal and spectral characteristics. 
 A quantitative evaluation of the method used is not always possible because mitigation algorithms are generally 
 non-linear processes that also affect the noise characteristics and the gain calibration. 
 The toxicity of the method used, i.e. the negative effect invoked on the data by the deployed method, and the amount 
 of data loss resulting from the method are other guiding factors for evaluating the various methods. 
 
 Multiple methods need to be applied to deal with a more general RFI environment. Because RFI characteristics 
 change after each mitigation step and with increasing integration of the data, the cumulative effect of RFI mitigation 
 at subsequent stages is not a linear sum of what each method can do, but rather the sum of what is practically 
 possible at each step. 
 
 The cost of computing hardware capabilities and digital applications at radio astronomy observatories are rapidly 
 changing parameters. Upgrades of existing observing facilities and newly constructed instrumentation provide ample 
 opportunity to implement and use automated RFI mitigation algorithms. 
 These changing capabilities also result in increased observing bandwidths, higher time resolution, and higher spectral 
 resolution. 
 These increasingly large data volumes will force the introduction of automated data reduction pipelines and automated 
 mitigation procedures. 
 While the traditional user community of radio observatories does not or only reluctantly accepts automated mitigation 
 implementations, they will be forced to do so in the future by the sheer volume of the data being handled. 
 
 New telecommunication and broadcasting technologies are reaching the market place, and many of these involve
 unlicensed mobile devices. 
 Their movement is difficult to control and they will rapidly affect observatory operations. 
 Algorithmic research is needed in order to eliminate the signals of these devices from the data. 
 In particular, spread spectrum (ultra-wide band) devices will pose problems for passive services because their digital 
 modulation schemes do not respect the boundaries of spectrum allocations. 
 Current estimates suggest that the number of transmitting devices used by each person is set to increase dramatically 
 and many of these devices will rely on dynamic spectrum access. 
The discovery space for radio astronomy is determined to a significant degree by the technical characteristics of the 
observing system and by limiting factors such as the RFI environment. 
While new generation instruments seek out the most pristine environments, existing facilities need to coexist with their 
local environment. 
In order to prevent the RFI environment becoming the limiting factor for each of the existing observatories, spectrum 
management, both internal and external, as a method to control this environment must remain a very high priority. 
Both observatory management and astronomers should be taking RFI issues seriously.


\end{document}